\documentclass[a4paper,fleqn,usenatbib]{mnras}

\usepackage{newtxtext,newtxmath}

\usepackage[T1]{fontenc}
\usepackage{ae,aecompl}
\usepackage{lipsum}                     
\usepackage[colorinlistoftodos]{todonotes}

\usepackage{graphicx}	
\usepackage{amsmath}	
\usepackage{amssymb}	

\title[MAGIC III: evidence of a dwarf stripping a dwarf]{The Magellanic Inter-Cloud Project (MAGIC) III: First spectroscopic evidence of a dwarf stripping a dwarf}

\author[R. Carrera et al.]{Ricardo Carrera,$^{1,2}$\thanks{E-mail: rcarrera@iac.es}
Blair C. Conn,$^{3}$
Noelia E. D. No\"el,$^{4}$
Justin I. Read,$^{4}$
\newauthor
\'Angel R L\'opez S\'anchez,$^{5,6}$
\\
$^{1}$Instituto de Astrof\'{\i}sica de Canarias, La Laguna E-3200, Tenerife, Spain\\
$^{2}$Departamento de Astrof\'{\i}sica, Universidad de La Laguna, La Laguna E-38205, Tenerife, Spain\\
$^{3}$Research School of Astronomy \& Astrophysics, Mount Stromlo
Observatory, Cotter Road, Weston Creek, ACT 2611, Australia\\
$^{4}$Department of Physics, University of Surrey, Guildford, GU2 7XH, UK\\
$^{5}$Australian Astronomical Observatory, P.O. Box 915, North Ryde, NSW 1670, Australia\\
$^{6}$Department of Physics and Astronomy, Macquarie University, NSW 2109, Australia
}

\date{Accepted XXX. Received YYY; in original form ZZZ}

\pubyear{2017} 

\begin{document}
\label{firstpage}
\pagerange{\pageref{firstpage}--\pageref{lastpage}}
\maketitle

\begin{abstract}
The Magellanic Bridge (MB) is a gaseous stream that links the Large (LMC) and Small (SMC) Magellanic Clouds. Current simulations suggest that the MB forms from a recent interaction between the Clouds. In this scenario, the MB should also have an associated stellar bridge formed by stars tidally stripped from the SMC by the LMC. There are several observational evidences for these stripped stars, from the presence of intermediate age populations in the MB and carbon stars, to the recent observation of an over-density of RR Lyrae stars offset from the MB. However, spectroscopic confirmation of stripped stars in the MB remains lacking. In this paper, we use medium resolution spectra to derive the radial velocities and metallicities of stars in two fields along the MB. We show from both their chemistry and kinematics that the bulk of these stars must have been tidally stripped from the SMC. This is the first spectroscopic evidence for a dwarf galaxy being tidally stripped by a larger dwarf.
\end{abstract}




\section{Introduction} \label{sec:sect1}

\noindent
The Magellanic Bridge (MB) was discovered by \citet{1963AuJPh..16..570H} as an \mbox{H\,{\sc i}} gaseous structure connecting the Small (SMC) and the Large (LMC) Magellanic Clouds. Two further streams of gas -- the Leading Arm and the Magellanic Stream -- were later discovered by \citet{1998Natur.394..752P} and \citet{1974ApJ...190..291M}, respectively. Initial attempts to reproduce the Magellanic System from dynamical simulations assumed multiple strong pericentric passages with the Milky Way. These were able to reproduce several features of the Magellanic System through the action of tidal stripping, ram pressure stripping, or a combination of the two \citep[see][for a recent compilation]{2016ARA&A..54..363D}.

However, recent measurements of the Magellanic Cloud's proper motions \citep[e.g.][]{2006ApJ...638..772K,2007MNRAS.381L..16B} have dramatically changed our view of the Magellanic system. Their large orbital velocities make the classical assumption of multiple encounters with the Milky Way unlikely. In this context, new hypotheses have emerged. \citet{2007ApJ...668..949B} introduced the `first infall' scenario. In this model, the Magellanic Clouds were bound to one another for a long period of time and are just now on their first infall to the Milky Way. In this picture, the MB and tail form as a result of a recent tidal interaction between the Clouds, prior to their recent accretion onto the Galaxy \citep{2012MNRAS.421.2109B}. An alternative scenario was proposed by \citet{2012ApJ...750...36D}. In this model, the LMC and the SMC have had multiple passages around the Galaxy, becoming a close binary pair only $\sim$2 Gyr ago. The Magellanic Stream and Leading Arm form from the first interaction between the Clouds, while the latest interaction $\sim$ 250 Myr ago formed the MB. The star formation histories of the LMC and SMC exhibit two correlated bursts of star formation at
$\sim$2 Gyr ago and $\sim$500 Myr ago that are consistent with such close interactions between the Clouds  \citep[e.g.][] {2007AJ....133.2037N,2009ApJ...705.1260N}.

Independently of the initial assumptions about the Magellanic Cloud's orbits, all of the latest models favour tidal stripping as the primary mechanism for forming the Leading and trailing arms and the MB \citep[e.g.][]{2014MNRAS.444.1759G}. Such models predict that, alongside the already observed gaseous structures, there must be companion stellar tidal debris comprised of intermediate-age and old, $\gtrapprox$ 1 Gyr, stars. What differentiates the different models are the number of stars stripped and their origin. For example, Model 2 from \citet{2012MNRAS.421.2109B} predicts a factor to $\sim$5 more stars in the MB as compared with their Model 1. In both cases, these stars originate in the SMC. By contrast, \citet{2014MNRAS.444.1759G} predict that the MB should contain stars stripped from both the SMC and LMC.

\begin{figure*}
\includegraphics[width=\textwidth]{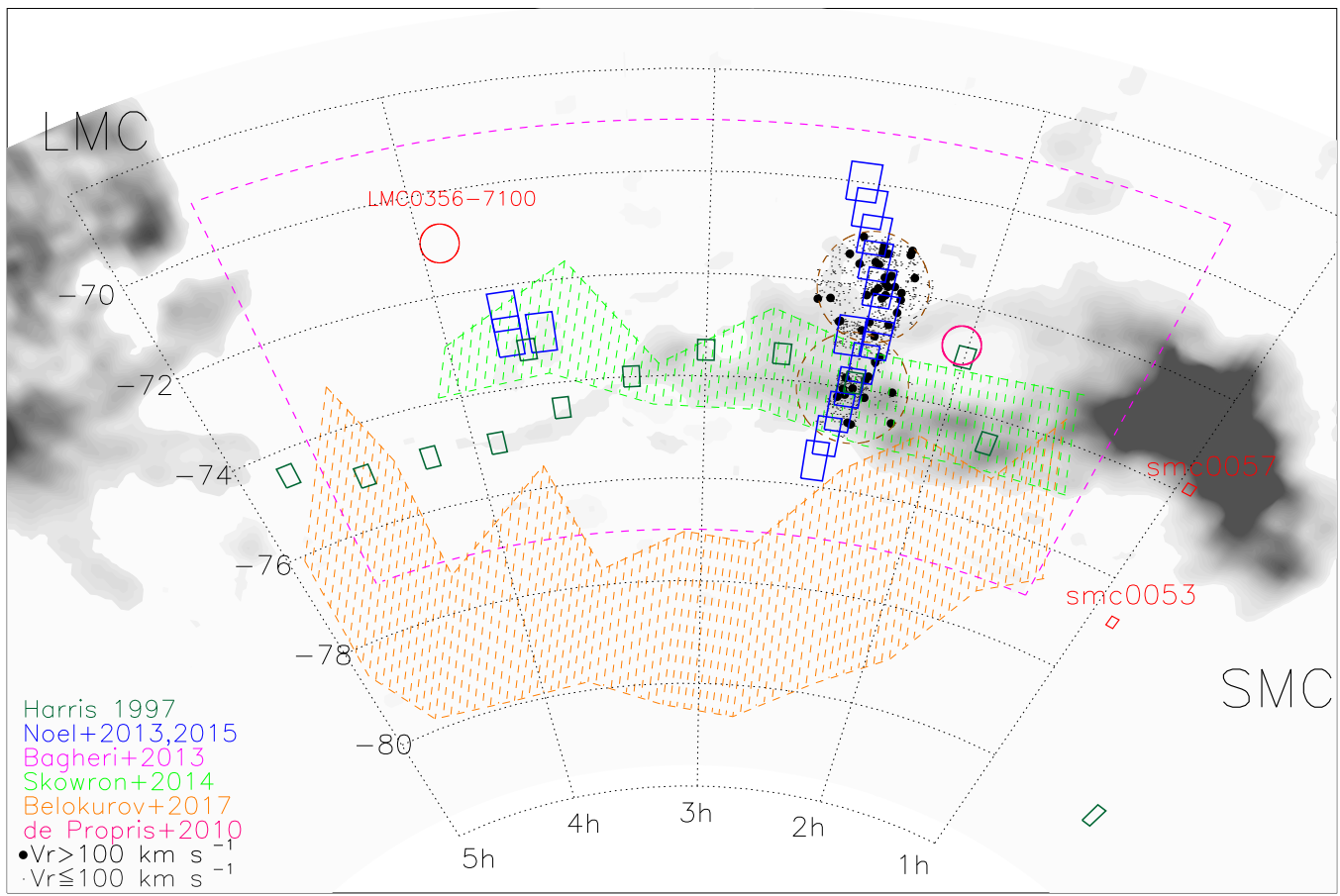}
\caption{The MB region in a Hammer equal area projection. The grayscale contours represent the \mbox{H\,{\sc i}} emission integrated over the velocity range 80 $\leq$V [km\,s$^{-1}$]$\leq$400, where each contour represents the \mbox{H\,{\sc i}} column density taken from the Leiden/Argentine/Bonn -- LAB -- survey of Galactic \mbox{H\,{\sc i}} \citep{2015A&A...578A..78K}. Regions studied by different authors have been plotted in different colors: \citet{2007ApJ...658..345H} in dark green; \citet{2010ApJ...714L.249D} in pink; and  \citet{2013ApJ...768..109N,2015MNRAS.452.4222N} in blue. The region analyzed by \citet{2013A&A...551A..78B} is denoted by dashed magenta lines. Green and orange dashed lines denote the young and RR Lyrae stars reported by \citet{2014ApJ...795..108S} and \citet{2016MNRAS.tmp.1588B}, respectively. The smc0053 and smc0057 fields studied by \citet{2008AJ....136.1039C} in the SMC and the field LMC0356-7100 analyzed by \citet{2011AJ....142...61C} in the LMC are marked in red. The grey dots and filled black circles are stars with radial velocities above and below 100 km s$^{-1}$ (the velocity cut used to separate MB stars from the Galactic foreground), respectively. \label{fig:spatial}}
\end{figure*}

The search for a stellar counterpart to the gaseous MB remains inconclusive. It is well known that there are young stars ($\lesssim$100 Myr) in this area, most likely formed in-situ \citep[e.g.][]{1990AJ.....99..191I}. In fact, \citet{2014ApJ...795..108S} have shown that these young stars form a continues stream linking both galaxies. However, intermediate-age and old stars have proven more elusive. \citet{1998AJ....115..154D} and \citet{2007ApJ...658..345H} were unable to find any in the MB region. However, more recent studies with deeper data find an excess of stars in the regions of the color-magnitude diagram (CMD) that are expected to be populated by these stars, e.g. the red clump, at the distance of the MB. \citet{2011ApJ...737...29O} found a distinct population in the LMC with a  median metallicity of [M/H]=-1.23 dex, substantially different from the bulk of the LMC disk population. This discrepancy in the metallicity lead Olsen et al. to claim that the kinematically distinct population found in the LMC was accreted  from the SMC. \citet{2011ApJ...733L..10N} reported a large azimuthally-symmetric metal-poor stellar population up to $\sim$11 kpc from the SMC centre that was well-fitted by an exponential profile. However, they were not able to determine whether these stars constituted a bound stellar halo or extra-tidal stars. \citet{2010ApJ...714L.249D} found a bimodal velocity distribution of the red giant branch (RGB) stars to the east and south of the SMC centre. The second peak at larger radial velocities was interpreted as SMC stars tidally striped from the LMC. However, this result has not been confirmed by recent studies \citep[e.g.][]{2014MNRAS.442.1663D}. A stellar structure in front of the SMC main body in the eastern region at a distance of 4\degr\ was reported by \citet{2013ApJ...779..145N}. They interpreted this as the tidally stripped stellar counterpart of the \mbox{H\,{\sc i}} gaseous bridge. As part of the MAGellanic Inter-Cloud project \citep[MAGIC;][]{2013ApJ...768..109N,2015MNRAS.452.4222N} we found a population with ages between 1 and 10\,Gyr with a spatial distribution more spread out than the younger stars. This older population has very similar properties to the stars located at $\sim$2\degr\ from the SMC centre suggesting than they were stripped from this region. \citet{2013A&A...551A..78B} found evidence for an older population in the MB that they suggest comprised tidal debris in the region 1$^{\rm h}$20$^{\rm m} \leq \alpha \leq$4$^{\rm h}$40$^{\rm m}$ and -69\degr$\geq \delta \geq$-77\degr. The radial density profiles of red clump stars in both galaxies, derived from the fourth phase of the Optical Gravitational Lensing Experiment (OGLE) by \citet{2014ApJ...795..108S}, shows a strong deviation. They suggest that this owes to the overlapping stellar halos of the Magellanic Clouds. Very recently, \citet{2016MNRAS.tmp.1588B} reported the existence of stellar tidal tail mapped with RR Lyrae in the Gaia DR1 database, not aligned with the gaseous MB and shifted by some $\sim$5\degr\ from the young main sequence bridge (see Fig.~\ref{fig:spatial}). Finally, a foreground population in the form of a distance bi-modality in the red clump distribution has been identified in the eastern SMC by \citet{2017MNRAS.tmp..202S} with the VISTA Magellanic Clouds (VMC) survey. The authors claim that the most likely explanation for this foreground population is tidal stripping from the SMC during its most recent encounter with the LMC. Thus, most of these studies support tidal stripping of stars from the SMC. However, the nature and origin of this stellar population in the inter-Cloud region is still not fully understood.

In this paper, we set out to unequivocally test the tidal origin of the MB stellar populations older than ${\sim}$1\,Gyr. To achieve this, we present the first spectroscopic analysis of the old stellar population in the MB region. If stars in the MB are indeed tidal debris, then they should have velocities and metallicities similar to those of the SMC and/or LMC stars. By determining whether the MB stars are more SMC-like or LMC-like, we will determine whether they were stripped primarily from the SMC, the LMC, or both.

This paper is organised as follows. In Section 2, we describe the target selection, observations, and data reduction. Radial velocities and stellar metallicity determination procedures are presented in Section 3. Finally, in Section 4 we discuss the implications of our results in the context of the Magellanic Clouds' dynamical evolution.

\section{Observational data}\label{sec:style}

\begin{table}
 \centering
\caption{Observed stars. The full version of this table is available in the online journal and at CDS.\label{tab:starsample}} 
 \resizebox{\columnwidth}{!}{
 \begin{tabular}{@{}lcccc@{}}
\hline
ID & $\alpha_{2000}$ & $\delta_{2000}$ & $K_\mathrm{S}$ & $V_{r}$ \\
   &   (deg)         &      (deg)      &   (mag)   &   (km s$^{-1}$)\\
   \hline
2M02251436-7205447 & 36.309831 & -72.0957544 & 13.685 & 59.3$\pm$2.9\\
2M02250915-7202337 & 36.288123 & -72.0426988 & 15.111 & 69.1$\pm$3.6\\
\hline
\end{tabular}}
\end{table}

\begin{figure}
\includegraphics[width=\columnwidth]{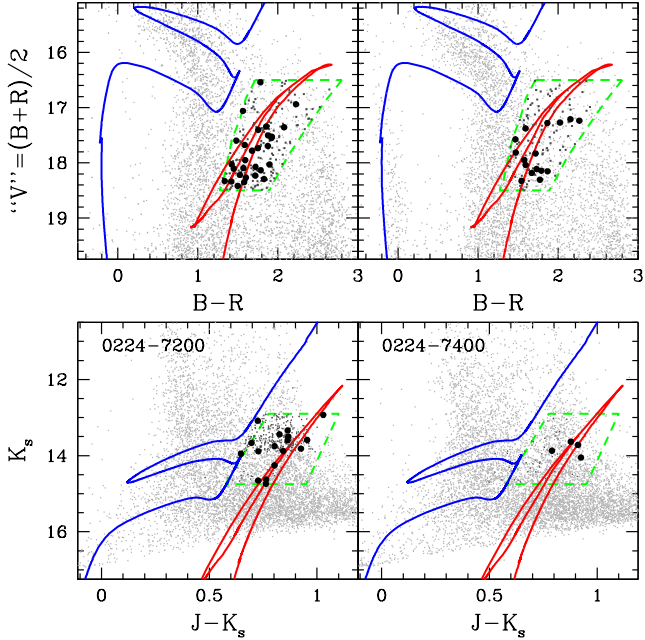}
\caption{Regions used to select the observed stars (green dashed lines) superimpoed on the CMDs of the 0224-7200 (left) and 0224-7400 (right) fields, respectively. Isochrones with age 100 Myr and metallicity Z = 0.004 (blue), and age 10 Gyr and metallicity Z = 0.002 (red) have been overplotted for reference. Light grey and black points show stars observed spectroscopicaly with radial velocities below and above 100 km s$^{-1}$ (the velocity cut used to separate MB stars from the Galactic foreground).\label{fig:fig1}}
\end{figure}

We have selected two fields studied in previous MAGIC papers centred at [$\alpha$ = $+02$\fh4, $\delta$ = $-72$\fdg0 (0224-7200)] and \mbox{[$\alpha$ = 02\fh4, $\delta$ = -74\fdg0 (0224-7400)],} respectively. The former is located at 6\fdg9 and 14\fdg1 from the SMC and LMC centres, respectively; the later at 6\fdg5 and 14\fdg0 from the SMC and LMC centres, respectively. The potential spectroscopic targets, selected from the expected position of the upper RGB in the CMDs, are shown in Fig.~\ref{fig:fig1} (large filled circles). This sample has been extended by selecting additional potential targets from the 2MASS \citep[Two Micron All Sky Survey; ][]{2006AJ....131.1163S}. To do that, we defined a region around the location of the objects previously selected from optical photometry (see Fig.~\ref{fig:fig1}). The observations were secured on the nights of the 12th and 13th November 2014 and 14th November 2016 with the AAOmega spectrograph \citep{2004SPIE.5492..389S} fed by the Two Degree Field (2dF) multi-object systems installed at the prime focus of the Anglo-Australian Telescope located at Siding Spring Observatory (Australia)\footnote{Program I.D.: ATAC/2014B/104, and S/2016B/04}. 2dF+AAOmega is a dual-beam spectrograph that allows to allocate up to 400 2"-size optical fibres within a 2\degr\ field of view. The blue arm was configured to observe blue targets located in the upper young main sequence, the analysis of which will be presented in a forthcoming paper. In the red arm, we used the grating 1700D centred on $\sim$8500\,\AA, providing a spectral resolution of $R\sim 8500$. Three different configurations were observed, two in a field at $+02$\fh4 and $-72$\fdg0 (0224-7400), for which we acquired three exposures of 3600~s for each of them, and one in a field at $+02$\fh4 and $-74$\fdg0 (0224-7400), for which we acquired a total of 6 exposures of 3600~s. The area covered by these fields is marked as dashed yellow circles in Fig.~\ref{fig:spatial}. In total, we obtained spectra with a signal-to-noise ratio larger than 5 for 514 stars listed in Table~\ref{tab:starsample}.

The initial steps of the data reduction, including bias subtraction, flat-field normalisation, fibre tracing and extraction, and wavelength calibration, were performed with the dedicated 2dF data reduction pipeline \citep[2dfdr\footnote{See https://www.aao.gov.au/science/software/2dfdr}; ][]{2010PASA...27...91S}. Our own software was used to subtract the emission sky lines following the procedure described in detail by \citet{carreraking1}. Briefly, the scale factor that minimises the sky line residuals is searched for by comparing the spectrum observed in each fibre with a master sky spectrum obtained by averaging the spectra of the nearest ten fibres placed on sky positions. After sky subtraction, the spectra are then normalised by fitting a low order polynomial. 

\section{Radial velocities and Metallicities}

\begin{figure}
\includegraphics[width=\columnwidth]{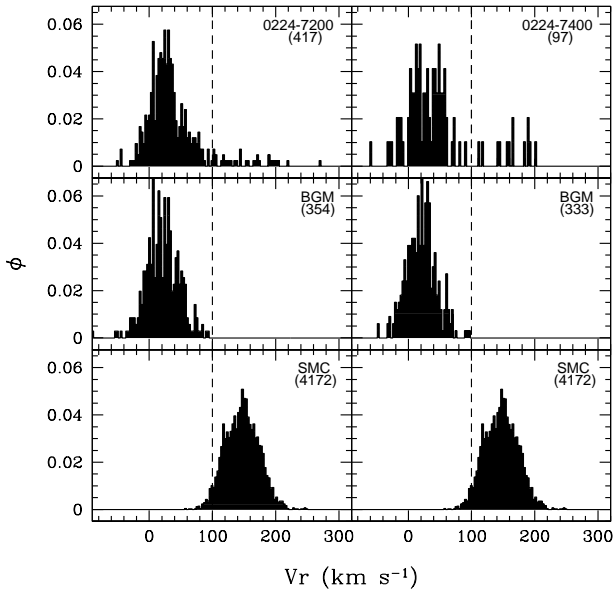}
\caption{Top: the velocity distributions of stars observed in fields 0224-7200 (left) and 0224-7400 (right). Middle: the distribution predicted by the BGM in each field. Bottom: the SMC velocity distribution from \citet{2014MNRAS.442.1663D}. Dashed lines mark the separation between Galaxy foreground and MB populations.\label{fig:distvr}}
\end{figure}

The radial velocities of the observed stars were calculated by comparing the observed spectra with a grid of synthetic spectra using the classical cross-correlation method. Details about the procedure and the grid of synthetic spectra used can be found in \citet{carreraking1}. In brief, the velocities are determined in three steps. (1) Each object spectrum is cross-correlated with a reference synthetic spectrum to obtain an initial shift for all of them. In this case, we chose: [M/H]=$-0.5$ dex; [$\alpha$/H]=$+0.0$ dex; $\xi$=1.5 km\,s$^{-1}$; T$_{\rm eff}$=4,500 k; and log $g$=2.0 dex. (2) After applying this initial shift, the observed spectrum is compared with the whole grid in order to identify the model parameters that best reproduce it through a $\chi^2$ minimisation using FER\reflectbox{R}E \citep{2006ApJ...636..804A}. (3) The best-fit synthetic spectrum is cross-correlated again with the observed spectrum in order to refine the shift between both. The heliocentric radial velocities derived for each star are listed in Table~\ref{tab:starsample}.

The velocity distribution of observed stars is shown in top panels of Fig.~\ref{fig:distvr}. There is a peak centred between 20 and 30 km\,s$^{-1}$ with a clear tail that extends from 100 to 200\,km\,s$^{-1}$. To better understand the obtained distribution we have plotted the prediction of the Besan\c{c}on Galaxy Model\footnote{Available at \url{http://model.obs-besancon.fr/}.} \citep[BGM;][]{2003A&A...409..523R} at the same position of the observed stars. This has been computed assuming the typical uncertainties for our radial velocities of 3\,km\,s$^{-1}$. We restricted our comparison to those stars located in the same region of the CMD as that of our target star locations, scaled to reproduce the height of the observed velocity distribution.

The BGM reproduces quite well the shape of the distribution between $\sim -10$ to $\sim$90 km\,s$^{-1}$. However, the model does not predict any star above 100 km\,s$^{-1}$. The velocity distribution of SMC stars, bottom panels of Fig.~\ref{fig:distvr}, is centred at 147.8$\pm$0.5 km\,s$^{-1}$ with a dispersion of 26.4$\pm$0.4 km\,s$^{-1}$ according to \citet{2014MNRAS.442.1663D}. This agrees with the tail at velocities larger than 100 km\,s$^{-1}$ of the observed distribution. In the case of the LMC, the radial velocities in the outskirts of the side that faces the SMC are typically larger than 200\,km\,s$^{-1}$ \citep[e.g.][]{2011AJ....142...61C}. This could explain the observed objects with the radial velocities around $\sim$200\,km\,s$^{-1}$ and one star with a velocity of $\sim$270\,km\,s$^{-1}$, but not the majority of stars between 100 and 200\,km\,s$^{-1}$.

\begin{figure}
\includegraphics[width=\columnwidth]{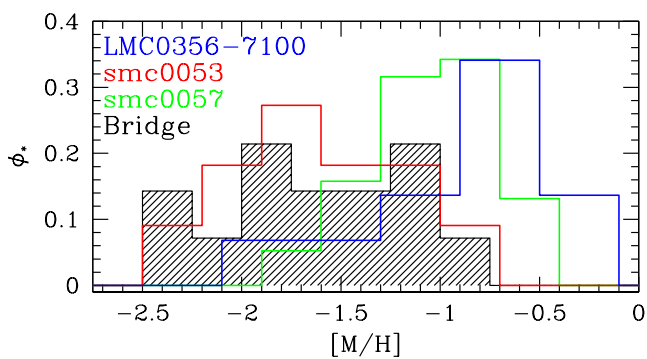}
\caption{The metallicity distribution of observed MB stars (shaded histogram). The metallicity distribution of an outer SMC field (red; smc0053, $\sim 4$\,deg from the SMC centre; \citealt{2008AJ....136.1039C}), an inner SMC field (green; smc0057, $\sim 1$\,deg from the SMC centre), and an LMC field (blue; LMC0356-7100; \citealt{2011AJ....142...61C}) have been over-plotted for reference (see Fig. \ref{fig:spatial} for the spatial location of these fields).\label{fig:feh}}
\end{figure}

We have complemented our analysis with the metallicities, [M/H], of observed stars obtained from an empirical relation between the strength of the \mbox{Ca\,{\sc ii}} triplet lines and a luminosity indicator, e.g. the absolute magnitude. We have used the relation obtained by \citet{2013MNRAS.434.1681C} using $M_{K_s}$ as a luminosity indicator. The absolute magnitude for each observed star was derived assuming a distance modulus of $(m-M)_0 = 18.6$ \citep{2016arXiv161201811W}, and using a reddening $E_{B-V}$ derived from the \citet{1998ApJ...500..525S} extinction maps. Finally, the strength of each line was determined by fitting its profile with a Gaussian plus a Lorentzian within a given bandpass following the procedure described by \citet{2007AJ....134.1298C}. The obtained metallicity distribution shown in Fig.~\ref{fig:feh} has 14 stars: those with V$_r\geq$100 km\,s$^{-1}$, signal-to-noise ratio larger than 10 and excluding the star with radial velocity $\sim$270 km\,s$^{-1}$. Our derived metallicities range from $-0.9$ to $-2.4$ dex with a median of $-1.68$ dex and a standard deviation of 0.47 dex. These values are similar to those obtained by \citet{2016arXiv161201811W} from RR Lyrae in the bridge, with an average of [M/H] = $-1.79$ dex.

For comparison, we have over-plotted in Fig.~\ref{fig:feh} the metallicity distribution of the innermost, smc0057 (green), and the outermost, smc0053 (red), fields studied by \citet{2008AJ....136.1039C} in the SMC (located on the right of Fig.~\ref{fig:spatial}). Clearly, the innermost field, located at a distance of $\sim 1$\degr\, from the SMC centre, is more metal rich than the bridge with a peak at $\sim -1$ dex. By contrast, the outermost field, at a distance of $\sim 4$\degr, has a distribution more similar to the bridge field, peaking at $-1.64$ dex. We have also over-plotted the metallicity distribution of the field LMC0356-7100 (blue) studied by \citet[][see Fig.~\ref{fig:spatial}]{2011AJ....142...61C}. This field is located at a distance of about $\sim 8$\degr\, and $\sim 14$\degr\, from the LMC and SMC centres, respectively. The metallicity distribution of this field is more metal rich than the outermost field studied in the SMC and in the MB, with a peak at $-0.87$ dex and reaching up to metallicities of -0.2 dex. We conclude that the stellar populations in the bridge are similar to the external parts of the SMC but more metal-poor than both the SMC centre and the outskirts of the LMC.

\begin{table}
 \centering
\caption{Average radial velocities, velocity dispersion, and the number of stars in each population.\label{tab:velo}} 
 \begin{tabular}{@{}lccc@{}}
\hline
Population & $\left\langle V_r\right\rangle $ & $\sigma$ & Stars \\
   & (km s$^{-1}$) & (km s$^{-1}$) & \\
   \hline
0224-7200 & 147.1 & 34.2 & 32\\
0224-7400 & 162.8 & 29.0 & 13\\
All & 151.7 & 33.3 & 45\\
\hline
$\leq$-1.85 & 109.4 & 9.6 & 4\\
-1.85 -- -1.2& 175.1 & 26.2 & 6\\
$\geq$-1.2 & 162.8 & 25.4 & 4\\
\hline
SMC & 147.8 & 26.4 & 4172\\
\hline
\end{tabular}
\end{table}

\section{Discussion}\label{sec:discussion}

\subsection{The SMC outskirts}

We have shown that stars older than ${\sim 1}$\,Gyr in the MB region, with radial velocities larger than 100 km\,s$^{-1}$, are unequivocally linked with the SMC. Except for one star with a velocity of 270\,km\,s$^{-1}$, we do not find objects with kinematics compatible with that of the LMC. This is in agreement with the results obtained by \citet{2014MNRAS.442.1663D} in the inner SMC, but contrasts with claims in \citet{2010ApJ...714L.249D} and dynamical simulations from \citet{2014MNRAS.444.1759G}. Note, however, that the stars analysed here are located almost 2\degr~ further from the SMC centre than those studied by \citet[][see Fig~\ref{fig:spatial}]{2010ApJ...714L.249D}.

\citet{2016AJ....152...58P} found a negative metallicity gradient in the inner SMC regions at $\leq$4\degr\ from its centre, and an inversion in the outer regions between 4\degr\ and 5\degr\ towards the Bridge. This can be interpreted as the presence of metal-rich stars in the outer regions stripped from the inner regions of the SMC. These results are in good agreement with the findings of \citet{2017MNRAS.tmp..202S} who used red clump stars. We do not find stars more metal-rich than $-0.9$\,dex and so we disfavour them coming from the inner $\sim$2\degr. From the synthetic CMD fitting technique, we conclude that the star formation histories of the fields studied here are similar to those obtained by \citet{2009ApJ...705.1260N} at a distance of $\sim$2.5\degr\ and by  \citet{2007ApJ...665L..23N} at $\sim$6\degr\ from the centre, both in the southern direction. However, the metallicity distributions obtained in the present work are more similar to the fields located at $\sim$4\degr\ in the same direction. This is in agreement with the stellar population with a shallow density profile reported by \citet{2011ApJ...733L..10N} at a radius between 3\degr\ and 7\fdg5. 

The shallow extended component of the SMC described above can be explained both by a bound stellar halo or by extra-tidal stars. To investigate further these two scenarios, we have computed the velocity dispersion of the stars with radial velocities larger than 100 km\,s$^{-1}$ (excluding the star with a velocity of 270\,km\,s$^{-1}$). There is no significant difference between the two fields studied (Table~\ref{tab:velo}). All together, these stars have an average velocity of 151.7\,km\,s$^{-1}$ and a dispersion of 33.3\,km\,s$^{-1}$. The velocity dispersion is in good agreement with the value obtained for the SMC $\sigma=26.4$\,km\,s$^{-1}$ by \citet{2014MNRAS.442.1663D} but the systemic velocity, $\left\langle V_r\right\rangle=147.8$\,km\,s$^{-1}$, is slightly lower than we find here for the inter-Cloud population. 

To further explore the above, we obtained the average velocity and dispersion in three metallicity bins: \mbox{[M/H]$< -1.85$;} $-1.85 < $[M/H]$< -1.2$; and [M/H] > -1.2, as listed in Table~\ref{tab:velo}. Since the SMC has a well defined age-metallicity relationship \citep[e.g.][]{2008AJ....136.1039C}, these three metallicity bins can also be thought of as a proxy for three age bins. We find that the most metal-poor stars have a mean radial velocity very different from the other stars, and a small dispersion of just 9.6\,km\,s$^{-1}$ (see Table~\ref{tab:velo}). This suggests that these owe to foreground contamination from Milky Way halo stars. For the other two more metal-rich bins, we find 
velocity dispersions in remarkable agreement with that of the SMC: 26.2\,km\,s$^{-1}$ and 25.4\,km\,s$^{-1}$, respectively (see Table~\ref{tab:velo}). The excellent agreement between the velocity dispersion of the more metal-rich inter-Cloud stars and that observed for stars in the SMC reinforces our hypothesis that these inter-Cloud stars originated in the SMC. 

The SMC has a stellar mass of $M_* {\sim} 4.6 \times 10^8$\,M$_\odot$ \citep{2012AJ....144....4M}. Thus, we expect from abundance matching that it would have inhabited a dark matter halo of mass $M_{200} {\sim} 7-9 \times 10^{10}$\,M$_\odot$ before infall \citep[e.g.][]{2017MNRAS.tmp..206R}, consistent with dynamical models of the SMC \citep{2009MNRAS.395..342B}. Similarly, the latest abundance matching and dynamical mass estimates for the LMC place it in a $M_{200} {\sim} 2 \times 10^{11}$\,M$_\odot$ halo before infall \citep{2016MNRAS.456L..54P}. Treating the Clouds as point masses moving on a pure radial orbit, and setting the tidal radius of the SMC to $r_t = 2.1$\,kpc (based on the similarity of the stars in the inter-Cloud region with stars $\sim$2\degr\ from the SMC centre), we can derive the pericentric radius of the recent SMC-LMC encounter that formed the MB as: $r_p \sim r_t\left(\frac{M_{\rm LMC}}{4 M_{\rm SMC}}\right)^{1/3} \sim 1.8$\,kpc \citep{2006MNRAS.366..429R}. This is in excellent agreement with recent models of tidal interactions between Clouds, where such a close encounter has been proposed to explain the LMC's off-centre stellar bar \citep{2012MNRAS.421.2109B}. This lends further support to a tidal origin for the inter-Cloud stars. 

\subsection{Comparison with simulations}

\begin{figure}
\includegraphics[width=\columnwidth]{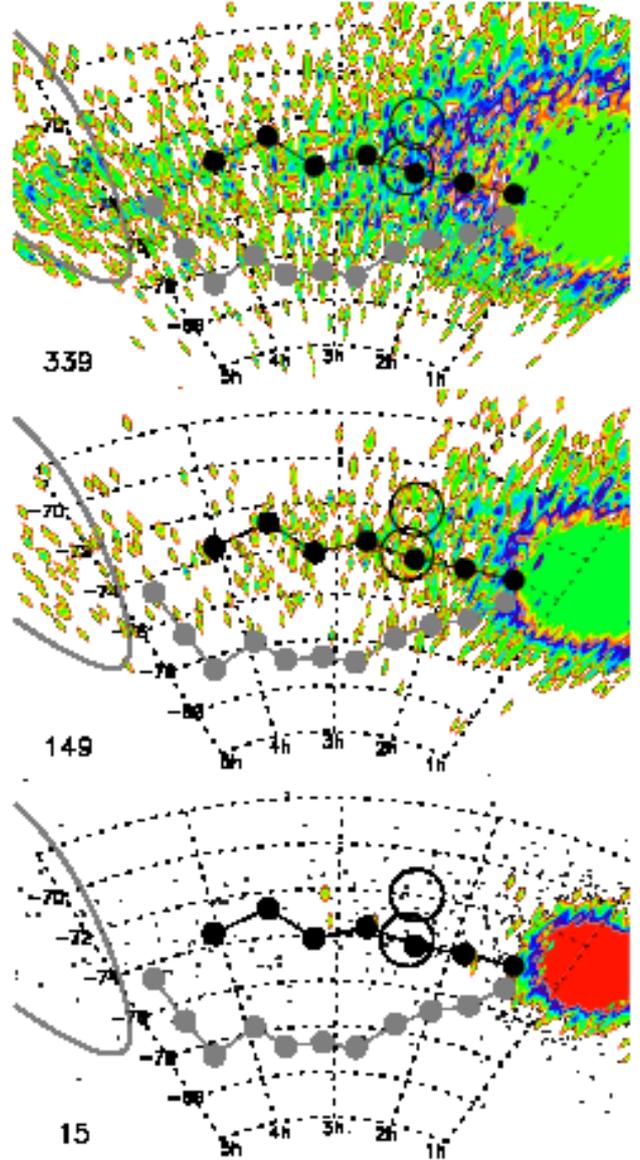}
\caption{The distribution of SMC stars predicted by the  \citet{2012ApJ...750...36D} ``model 1" (top), ``model 2" (middle) and ``model 3" (bottom) in the inter-Cloud region. Contours show the particle counts. In the bottom panel, the dots represent individual star particles that have been omitted in the other panels for clarity. The ellipse on the left shows the position of the LMC. The filled black and gray circles mark the location of the young and RR Lyrae stellar bridges. The open black circles mark the locations of the two fields observed in this paper. The number of stars predicted by each model in the location of our observed fields (circles) is listed in bottom-left corner.\label{fig:diazbekki}}
\end{figure}

The kinematics, photometry and chemistry of stars in the MB all point to them having been tidally stripped from the SMC. In this section we perform a more direct comparison with the dynamical simulations from \citet{2012ApJ...750...36D}. In particular, we compare the location of tidally stripped stars in the MB region of the simulations with those in the locations of the fields sampled in this paper.

Although the \citet{2012ApJ...750...36D} simulations are based on the ``bound scenario", in their favoured model the LMC and SMC suffer strong tidal interactions only recently, with two close passages $\sim$2 Gyr and $\sim$250 Myr ago. Portions of the gaseous disk of the SMC are stripped away during each of these strong encounters. In particular, the most recent one forms the gaseous bridge.

Since the gravitational field of the LMC acts in the same manner on gas and stars, a tail of stars pulled out from the SMC into the MB is expected. \citet{2012ApJ...750...36D} assume a multi-component SMC, consisting of a spherical dark matter halo, central disc and a more extended spheroidal component. They generated three different models where the size of the spheroid was varied: ``model 1" with a Plummer scale radius of 7.5\,kpc; ``model 2" with 5.0\,kpc; and ``model 3" with 2.5\,kpc, respectively. This is because it is still unclear how stellar populations of different ages and metallicities are distributed within the tidal radius of the SMC. The top panel of Fig~\ref{fig:diazbekki} shows the predicted distribution of stars for each model. It is clear that stripped stars in all three cases are found in the MB region. However, the number of stars stripped in model 3 is much lower than for the other two models, owing to its more concentrated spheroidal population. Another interesting result is that in all cases, stripped SMC stars are captured by the LMC. This is in agreement with the peculiar stellar populations reported in the LMC by \citet{2011ApJ...737...29O}. In all three models, the SMC maintains an approximately spheroidal outer population after the interaction with the LMC. However, clear tidal distortions are observed towards the LMC in models 1 and 2. The position of these tails agree relatively well with location of the observed gaseous bridge (black connected dots). On contrast, the models do not predict a significant number of tidally stripped object in the position of the recently discovered RR Lyrae bridge (gray connected dots). Although model 3 remains dense and compact, stripped stars in the MB are still present. In model 1, the edge of the spheroid almost reaches the position of our fields (open black circles). In this case, the populations there would be a mix of both bound and unbound stars. By contrast, models 2 and 3 predict mainly tidally stripped stars from the SMC at the positions of our observed fields. This is not a particularity of the \citet{2012ApJ...750...36D} simulations. All recent models based on the latest knowledge of the Magellanic System also predict the existence of tidally stripped stars in the inter-Cloud region to a greater or lesser extent \citep{2012MNRAS.421.2109B,2014MNRAS.444.1759G,2016MNRAS.tmp.1588B}.

\begin{figure}
\includegraphics[width=\columnwidth]{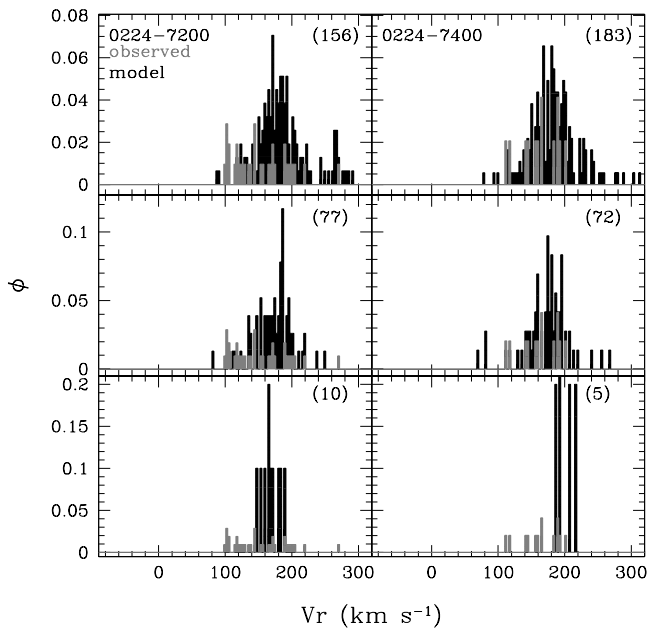}
\caption{Radial velocity distributions predicted by ``model 1" (top), ``model 2" (middle) and ``model 3" (bottom) in the 0224-7200 (left) and 0224-7400 (right) fields, respectively. The number of stars in each field is listed in top right corner of each panel. The velocity distribution of observed stars with velocities larger than 100 km\,s$^{-1}$  have been plotted as comparison (gray histograms). \label{fig:diazbekki_vr}}
\end{figure}

These simulations also provide valuable information about the expected radial velocities of the model stars at the positions of our fields. The velocity distributions predicted for each model at the locations of fields 0224-7200 and 0224-7400 are shown in Fig.\ref{fig:diazbekki_vr}. In the three models, the bulk of stars have radial velocities between $\sim$90 and $\sim$200 km s$^{-1}$. However, model 1 predicts a continuum tail towards $\sim$300 km s$^{-1}$ that is not observed in our fields. By contrast, model 2 only predicts a few stars between 200 and 300 km s$^{-1}$ while model 3 does not predict any. Our kinematic results favour models 2 and 3 over model 1. However, the spatial distributions observed in the SMC periphery favour models 1 and 2 over model 3 \citep[and see also][for a detailed discussion]{2012ApJ...750...36D}. Thus, taking the spatial and velocity data together, model 2 is closest to our observations.

\subsection{Dwarf/dwarf stripping}

The tidal stripping of satellite galaxies is expected to be a common process \citep[e.g.][]{2006MNRAS.366..429R,2014ApJ...794..115D}. Indeed, we see direct evidence for such stripping events in the Milky Way \citep[e.g.][]{1994Natur.370..194I}, M31 \citep[e.g][]{2001Natur.412...49I}, and other spiral galaxies in the Local Volume \citep[e.g.][]{2010AJ....140..962M,2012ApJ...748L..24M}. Being the most numerous type of galaxies, interactions between dwarf systems have been widely reported \citep[e.g.][]{2008A&A...491..131L,2009A&A...508..615L}. In fact, stellar tidal tails have been associated with the interaction of the Magellanic analog system formed by NGC~4485 and NGC~4490 \citet{1998AJ....115.1433E}. The findings shown in this work confirm for the first time the existence of a stellar population older than 1 Gyr in the Magellanic inter-Cloud area, unequivocally related to the SMC. Moreover, these stars have a metallicity distribution similar to that of the SMC outskirts, as shown in Fig. \ref{fig:feh}. This represents the first spectroscopic evidence for dwarf-dwarf stripping in the Universe.

\section{Conclusions}\label{sec:conclusions}

We have used medium resolution spectra for 514 red giant stars in the Magellanic Bridge (MB). Our key findings are as follows:

\begin{itemize}
\item The chemistry and kinematics for 39 of the target stars are consistent with those located $\sim 2$\,deg from the centre of the SMC, but inconsistent with LMC stars. We conclude, therefore, that these stars were tidally stripped from the SMC.

\item We used the above to estimate the tidal radius of the SMC, finding $r_t \sim 2.1$\,kpc. Using the latest estimates of the pre-infall masses of the LMC and SMC, we then estimated the closest passage between the Clouds to be $r_p \sim 1.8$\,kpc. Such a close encounter has been invoked to explain the LMC's off-centre stellar bar \citep{2012MNRAS.421.2109B}.

\item We compared the spatial location and kinematics of stars in the MB region to the simulations from \citet{2012ApJ...750...36D}. We found that their ``model 2'' provided the best qualitative match to our data. In this model, in addition to a stellar disc and dark matter halo, the SMC has an outer spheroidal population with a Plummer scale length of 5\,kpc. Models with a more concentrated spheroid produce too little tidal debris in the MB, while those with a more extended spheroid produce a tail to large radial velocities that is not observed.

\item Our results represent the first spectroscopic evidence for a dwarf galaxy being tidally stripped by a larger dwarf.
\end{itemize}

\section*{Acknowledgements}

We would like to thank the anonymous referee for their helpful comments that have improved this paper. We would also like to thank J. Diaz and K. Bekki for providing us with their $N$-body models. The research leading to these results has received funding from the European Community's Seventh Framework Programme (FP7/2013-2016) under grant agreement number 312430 (OPTICON). BCC acknowledges the support of the Australian Research Council through Discovery project DP150100862. This work was supported by the Spanish Ministry of Economy through grant AYA2014-56795P.







\bsp	
\label{lastpage}
\end{document}